\documentclass[structabstract]{aa}
\usepackage{graphicx}
\usepackage{txfonts}
\usepackage{natbib}
 
\begin{document}

\title {Modelling the chemical evolution of the Galaxy halo}

\author {G. Brusadin\inst{1} \thanks {email to: brusadin@oats.inaf.it} \and F. Matteucci\inst{1,2,3}\and D. Romano \inst{4}} 
 \institute{Dipartimento di Fisica, Sezione di Astronomia, Universit\`a di Trieste, via G.B. Tiepolo 11, I-34131, Trieste, Italy 
\and INAF, Osservatorio Astronomico di Trieste, via G.B. Tiepolo 11, I-34131, Trieste, Italy
\and INFN, Sezione di Trieste, Via Valerio 2, I-34134, Trieste, Italy
\and INAF, Osservatorio Astronomico di Bologna, Via Ranzani 1, I-40127 Bologna, Italy}

\date{Received xxxx / Accepted xxxx}

\abstract 
{The stellar halo provides precious information about the Galaxy in its early stages of evolution because the most metal-poor (oldest) stars in the Milky Way are found there.}
{We study the chemical evolution and formation of the Galactic halo through the analysis of its stellar metallicity distribution function and some key elemental abundance patterns. We also test the effects of a possible population III of zero-metal stars.}
{Starting from the two-infall model for the Galaxy, which predicts too few low-metallicity stars, we add a gas outflow during the halo phase with a rate proportional to the star formation rate through a free parameter, $\lambda$. In addition, we consider a first generation of massive zero-metal stars in this two-infall + outflow model adopting two different top-heavy initial mass functions and specific population III yields.}
{The metallicity distribution function of halo stars, as predicted by the two-infall + outflow model shows a good agreement with observations, when the parameter $\lambda=14$ and the time scale for the first infall, out of which the halo formed, is not longer than 0.2 Gyr, a lower value than suggested previously. Moreover, the abundance patterns [X/Fe] vs. [Fe/H] for C, N and $\alpha$-elements O, Mg, Si, S, Ca show a good agreement with the observational data, as in the case of the two-infall model without outflow. If population III stars are included, under the assumption of different initial mass functions, the overall agreement of the predicted stellar metallicity distribution function with observational data is poorer than in the case of the two-infall + outflow model without population III.}
{We conclude that it is fundamental to include both a gas infall and outflow during the halo formation to explain the observed halo metallicity distribution function, in the framework of a model assuming that the stars in the inner halo formed mostly in situ. Moreover, we find that it does not exist a satisfactory initial mass function for population III stars which reproduces the observed halo metallicity distribution function. As a consequence, there is no need for a first generation of only massive stars to explain the evolution of the Galactic halo.}

\keywords{Galaxy: abundances - Galaxy: evolution - Galaxy: halo}

\titlerunning{   }
\authorrunning{  }
\maketitle
 
\section{Introduction}

The stellar halo of the Milky Way contains useful information about the earliest phases of Galactic evolution, and galaxy evolution in general, because the most metal-poor and oldest stars known are found there (Bessel \& Norris 1984; Christlieb et al. 2002; Frebel et al. 2005; Aoki et al. 2006; Norris et al. 2007; Caffau et al. 2011). These stars are the local, accessible counterparts of unreachable high-redshift objects. As such, they constitute a unique test bed of theories of first star formation, evolution and nucleosynthesis, as well as structure formation in the early Universe. In fact, as first shown by Eggen, Lynden-Bell \& Sandage (1962), crucial information on the formation and evolution of a stellar system can be obtained from the kinematics and chemical composition of its stars. In particular, low-mass stars ($m<1$ $\mathrm{M}_{\odot}$) live for a time comparable or much longer than the present age of the Universe retaining in their atmospheres a record of the chemical composition of the gas out of which they were born. Low-mass, metal-poor halo stars are thus fossils, whose chemical abundances contain information about the early stages of Galactic evolution, as well as the nucleosynthesis processes in the first generations of stars. The surface abundances of particular elements such as C, N and, possibly, O, however, could have been significantly altered by internal `mixing' processes or external influences, such as binarity or accretion of interstellar material (e.g. Spite et al. 2005; Frebel \& Norris 2011). Therefore, caution should be exercised when comparing model predictions to abundance data for those elements; heavier elements are cleaner probes of the chemical enrichment history of galaxies.
Traditionally, two models, introduced in the 1960s and 1970s, respectively, have been used to describe the evolution of the Galactic halo. In the monolithic scenario originally proposed by Eggen, Lynden-Bell \& Sandage (1962), a fast dissipative collapse on a time scale of only a few times $10^8$ years is envisaged, during which massive stars die and enrich the gas with heavy elements. The second approach was proposed by Searle \& Zinn (1978). On the basis of the lack of an abundance gradient for the outer halo globular clusters (GCs), these authors suggested that \emph{`these clusters formed in a number of small protogalaxies that subsequently merged to form the present Galactic halo'} on a time scale $>$1 Gyr. This model is more similar to the currently popular hierarchical paradigm for structure formation. According to the currently favoured $\Lambda$ Cold Dark Matter ($\Lambda$CDM) cosmological model, in fact, large galaxies formed from the hierarchical aggregation of smaller subunits, that collapsed first (e.g. White \& Rees 1978; White \& Frenk 1991). In this framework, different processes ---infall, mergers and interactions with other galaxies--- contribute to different extents to define the properties of galaxies, relative to the local environment.
On the observational side, there is mounting evidence for a complex halo structure: an inner/outer halo dichotomy has been observed, with the two substructures likely having a different origin (Carollo et al. 2007), and a wealth of stellar streams and coherent substructures are identified in the Galactic halo (e.g. Ibata, Gilmore \& Irwin 1995; Belokurov et al. 2006; Starkenburg et al. 2009), as well as in the stellar haloes of external galaxies (e.g. Ferguson et al. 2002; Irwin et al. 2005; de Jong et al. 2007; Mouhcine et al. 2010; Ibata et al. 2011). In addition, the slope of the halo density profile in the Galaxy seems to show a break beyond $R \sim$25 kpc and the inner halo regions are flattened (Sesar et al. 2011; Deason et al. 2011). Last but not least, it is now agreed that in order to explain the relative fraction of chemically peculiar-to-normal stars in Galactic GCs a considerable fraction of the halo (at least half) must have originated in proto-GCs (Gratton \& Carretta 2010, and references therein): these  proto-GCs could be the protogalactic fragments invoked by Searle \& Zinn (1978). Mackey et al. (2010), using a sample of newly-discovered GCs from the Pan-Andromeda Archaeological Survey (PAndAS), are able to associate the GCs to large, coherent stream features in the M31 halo, thus providing clearcut evidence that the outer halo GC populations in some galaxies are largely accreted. While it is likely that the \emph{outer} Galactic halo formed primarily by accretion/mergers (e.g. Zolotov et al. 2009; Tissera et al. 2012), we feel that the issue of the \emph{inner} halo formation ---did it form mainly through mergers of smaller systems or through a rapid, dissipational gas collapse?--- is still uncomfortably unsettled.

In the last decade several authors have modeled the formation of stellar haloes in a full cosmological framework, either by means of numerical simulations or using semi-analythical models (e.g. Bekki \& Chiba 2001; Scannapieco \& Broadhurst 2001; Abadi et al. 2006; Font et al. 2006, 2011; Tumlinson 2006; Salvadori et al. 2007; Zolotov et al. 2009; Brook et al. 2012, to name a few), and provided different answers to the above question. According to some simulations (e.g. Bullock \& Johnston 2005; Johnston et al. 2008; Cooper et al. 2010) the stellar haloes formed primarily through satellite accretion and disruption. However, models that form haloes by non-dissipative accretion and mergers are not able to produce significant metallicity gradients (e.g. De Lucia \& Helmi 2008; Cooper et al. 2010), at variance with observations (Carollo et al. 2007). Hence, there have been claims that a significant fraction of the halo stars must form in situ. For instance, from their simulations Tissera et al. (2012) find that inner spheroids have old, metal-rich, $\alpha$-enhanced stars which have formed primarily in situ, more than 40\% from material recycled through earlier galactic winds. Few accreted stars are found in the inner spheroid unless a major merger occurred recently. Font et al. (2011) find that, on average, 2/3 of the mass of the stellar spheroid form in situ; the large contribution of in situ star formation in the inner regions leads to the development of large-scale metallicity gradients. A two-phase galaxy formation is suggested by Oser et al. (2010): an early, rapid in situ star formation phase ---resembling in many aspects a monolithic collapse, though in the current paradigm the dissipative component would have formed as a result of gas-rich mergers at early times, rather than through a smooth gas accretion--- labelled `cold flow-driven' star formation (e.g. Dekel \& Birnboim 2006) is followed by a late merger-dominated period. Stars formed within the virial radius of the final system are classified as stars born in situ, while those formed outside it are the accreted stars.

In previous works, the metallicity distribution function (MDF) of long-lived 
stars has been most often the only chemical property of the final stellar 
systems used to constrain the models/simulations. Seldom, the trends of 
$\alpha$-elements-to-iron as a function of metallicity ([Fe/H]) have 
been used to this purpose.
In particular, in many hierarchical models of galaxy formation it is assumed the instantaneous recycling approximation which allows to compute only the evolution of the global metallicity (Z), dominated by oxygen. The observational data instead refer to the Fe abundance. Therefore, in order to compare model results and data one is forced to adopt a conversion between Z and Fe (see De Lucia \& Helmi 2008). In other papers, the Fe evolution was considered but no Type Ia supernovae were included (e.g. Salvadori et al. 2007), under the assumption that they do not occur in the halo. This is not true, since the very first Type Ia SNe can occur after 35 Myr, the lifetime of a 8~M$_{\odot}$ star (see Matteucci 2012). 
In this paper, we intend to study the stellar abundances and abundance ratios in great detail, in order to put constraints on the nature of Galactic halo stars through the chemistry.
Although our model is not explicitely motivated by the hierarchical theory of galaxy formation, our results can be considered complementary to the cosmological ones.
The discrepancy between the abundance patterns of surviving Local Group dwarf spheroidal galaxies (dSphs) and those of Milky Way halo stars has been quoted as evidence against the hierarchical structure assembly predictions (e.g. Venn et al. 2004). However, it was soon realized that most of the accretion events must have 
happened quite early on, at look-back times of $\sim$10 Gyr, before a significant chemical evolution in the accreted systems had taken place. Thus, the basic galactic building blocks are not expected to resemble any of the surviving dwarf galaxies, as far as their chemical properties are concerned (see, e.g., Robertson et al. 2005). 
Recent observational studies aiming at unraveling the origin of halo stars through a detailed analysis of their chemical abundances revealed the existence of two distinct halo populations with a systematic difference in [$\alpha$/Fe], [Cu/Fe], [Zn/Fe] and [Ba/Y], but not in [Mn/Fe], at a given metallicity (Nissen \& Schuster 2010, 2011, and references therein; see also Ishigaki et al. 2012). A dual distribution of [$\alpha$/Fe] is obtained in the simulations of Zolotov et al. (2009). These findings suggest that the inner halo formed at a different speed than the outer halo, and that this latter could have formed on a longer timescale or that the stars in the outer halo have been accreted.
Abundance ratios can be useful tools to understand whether the halo stars have been accreted or have formed in situ. First of all, if the majority of halo stars have been accreted from dwarf galaxies, one would expect a large spread in the [X/Fe] ratios. Such a big spread is not observed in the [$\alpha$/Fe] ratios but is present in [Ba, Eu/Fe] ratios. In general, the [X/Fe] vs. [Fe/H] trends for $\alpha$-elements in dSphs are different than those in the Milky Way stars. It has been shown (e.g. Lanfranchi \& Matteucci 2004) that dwarf spheroidals should have formed stars at a much lower rate than in the Galactic halo 
so to explain the more rapid decrease of the [$\alpha$/Fe] ratios with increasing [Fe/H] than in the halo and disk stars. However, some metal-poor stars in dwarf spheroidals ([Fe/H]$<-2.0$ dex) show [$\alpha$/Fe] ratios similar to halo stars. Prantzos (2008) compared the metallicity distribution functions, but not the [$\alpha$/Fe] ratios, of the Milky Way halo stars and its satellites and concluded that accretion of galaxies similar (but not identical) to the progenitors of present-day dwarf satellites of the Milky Way may well have formed the Galactic halo. Tumlinson (2006) also reproduced the MDF of the halo stars by means of a chemical model grounded in the hierarchical theory of galaxy formation.

In this work, we study the chemical evolution and formation of the Galactic halo by comparing the predictions of a chemical evolution model to the observational data of halo stars. In particular, we adopt the two-infall model of Chiappini et al. (1997) which assumes that the Galaxy formed by means of two main gas accretion episodes: during the first one the halo and the thick disk formed and during the second one the thin disk was assembled. The novelty here is that we include in the two-infall model a gas outflow during the halo formation phase. We then compare the model results to the MDF of the halo and to some elemental abundance patterns including disk stars. Previous papers (i.e. Hartwick 1976; Prantzos 2003; Cescutti \& Chiappini 2010) have already suggested a gas outflow from the halo but either this hypothesis was tested in very simple models or not tested against all the observational constraints (see discussion above). Furthermore, we test on our model the effects of a possible population III of first massive stars with zero-metal content. We compare our results with previous findings in the literature, and show that the issue of popIII stars is far from being settled. The other novelty of this paper is that we adopt the most recent stellar yields for massive and intermediate mass stars (Romano et al. 2010).

The paper is organized as follows: in Sect. 2 we describe the basic ingredients for galactic chemical evolution models and the main features of the two-infall model we started from; in Sect. 3 the two-infall + outflow model and its comparison with observations are presented. In Sect. 4 we test the effects of a population III on our model results and finally, in Sect. 5, we draw our conclusions.

\section{Modeling galactic chemical evolution: the two-infall model}

Our aim is to study the chemical evolution of the Galactic halo starting from the numerical two-infall model for the Galaxy formation developed first in Chiappini et al. (1997, 2001) and updated by Romano et al. (2010). The model predicts the mass abundances of H, D, \element[ ][3]{He}, \element[ ][4]{He}, \element[ ][7]{Li}, \element[ ][12]{C}, \element[ ][13]{C}, \element[ ][14]{N}, \element[ ][15]{N}, \element[ ][16]{O}, \element[ ][17]{O}, \element[ ][18]{O}, \element[ ][19]{F}, \element[ ][23]{Na}, \element[ ][24]{Mg}, \element[ ][27]{Al}, \element[ ][28]{Si}, \element[ ][32]{S}, \element[ ][39]{K}, \element[ ][40]{Ca}, \element[ ][45]{Sc}, \element[ ][46]{Ti}, \element[ ][51]{V}, \element[ ][52]{Cr}, \element[ ][55]{Mn}, \element[ ][56]{Fe}, \element[ ][58]{Ni}, \element[ ][59]{Co}, \element[ ][63]{Cu} and \element[ ][64]{Zn}, together with the gas density, the total mass density, the stellar density and the rate of Type Ia and Type II supernovae as functions of time and Galactocentric distance. 

\subsection{Two-infall model assumptions}

The two-infall model assumes that the Galaxy forms out of two main accretion episodes, almost completely dissociated. During the first episode, the primordial gas collapses very quickly and forms the spheroidal components (halo and bulge); during the second episode, the thin disk forms, mainly by accretion of a significant fraction of matter of primordial chemical composition plus traces of halo gas. The disk is built-up in the framework of the `inside-out' scenario of Galaxy formation (Larson 1976) which ensures the formation of abundance gradients along the disk (Matteucci \& Fran\c{c}ois 1989).

The Galactic disk is approximated by several indipendent rings 2 kpc wide, without exchange of matter between them. The rate of matter accretion in each shell is given by:

\begin{equation}
\frac{d\sigma_{I}(R,t)}{dt}=A(R)e^{-\frac{t}{\tau_H}}+B(R)e^{-\frac{t-t_{max}}{\tau_D}}
\end{equation}

where $\sigma_{I}(R,t)$ is the surface mass density of the infalling material, which is assumed to have primordial chemical composition; $t_{max}$ is the time of maximum gas accretion onto the disk, coincident with the end of the halo/thick disk phase and set here equal to 1 Gyr; $\tau_H$ and $\tau_D$ are the time scales for mass accretion onto the halo/thick disk and thin disk components, respectively. In particular, $\tau_H=0.8$ Gyr and, according to the `inside-out' scenario, $\tau_D(R)=1.033 \mathrm{(R/kpc)}-1.267$ Gyr (see Romano et al. 2000). The quantities $A(R)$ and $B(R)$ are derived from the condition of reproducing the current total surface mass density distribution in the halo and along the disk, respectively.

The SFR is (see Chiappini et al. 1997 for details):

\begin{equation}
\psi(R,t)=\nu(t)\Bigl(\frac{\sigma(R,t)}{\sigma(R_{\odot},t)}\Bigr)^{2(k-1)}\Bigl(\frac{\sigma(R,t_{Gal})}{\sigma(R,t)}\Bigr)^{k-1}\sigma_{gas}^k(R,t)
\end{equation}
where $\nu(t)$ is the star formation efficiency, $\sigma(R,t)$ is the total surface mass density at radius $R$ and time $t$ and $\sigma_{gas}(R,t)$ is the surface gas density, with the exponent $k=1.5$ given by the fit to observational constraints in the solar vicinity. The present surface mass density distribution of the disk is an exponential function of the scale lenght $R_D=3.5$ kpc normalized to $\sigma_D(R_{\odot}, t_{Gal})=54$ $\mathrm{M}_{\odot}$$\mathrm{pc}^{-2}$ (see Romano et al. 2000 for a discussion about the choice of these parameters). 
The total halo mass density profile is constant and equal to $17$ $\mathrm{M}_{\odot}$$\mathrm{pc}^{-2}$ for $R\le 8$ kpc and decreases as $R^{-1}$ outwards. In practice, we adopt the halo mass profile given by Binney \& Tremaine (1987): the density is proportional to $R^{-2}$ and this implies a surface mass density profile proportional to $R^{-1}$. The halo surface mass density at solar position is quite uncertain and we assumed it to be 17 $\mathrm{M}_{\odot}$$\mathrm{pc}^{-2}$ because the total surface mass density in the solar vicinity is $\sim$71 $\mathrm{M}_{\odot}$$\mathrm{pc}^{-2}$ (Kuijken \& Gilmore 1991) and $\sim$54 $\mathrm{M}_{\odot}$$\mathrm{pc}^{-2}$ corresponds to the disk surface mass density. The star formation efficiency is set to $\nu=2$ $\mathrm{Gyr}^{-1}$ in the halo and $\nu=1$ $\mathrm{Gyr}^{-1}$ in the disk to ensure the best fit to the observational features in the solar vicinity and becomes zero when the surface gas density drops below a certain critical threshold value.

Finally, the initial mass function (IMF) is the three-slope 
Kroupa et al. (1993) one
for the mass range 0.1-100 $\mathrm{M}_{\odot}$:
\begin{equation}
\left\{
\begin{array}{ll}
x=0.3 & \mathrm{if}~M\le0.5 \mathrm{M}_{\odot} \\
x=1.2 & \mathrm{if}~0.5<\frac{M}{\mathrm{M}_{\odot}}\le1.0\\
x=1.7 & \mathrm{if}~M>1.0 \mathrm{M}_{\odot}.
\end{array}
\right. 
\end{equation}

\subsection{Threshold gas density for star formation}

The gas density thresholds adopted in this work are 4 and 7 $\mathrm{M}_{\odot}$$\mathrm{pc}^{-2}$ during the halo and disk phase, respectively. These values were used also in Chiappini et al. (2001).
It is worth recalling that a surface density threshold for star formation has been observed in a variety of objects, including normal spirals, starburst galaxies and low surface brightness galaxies (e.g. Kennicutt 1989, 1998). The existence of a threshold in spheroidal systems (halos, bulges and ellipticals) is not yet clear although there are theoretical arguments by Elmegreen (1999) suggesting the existence of a threshold also in such systems. 
One of the main effects of the threshold in the Chiappini et al. (2001) model was that it naturally produced a hiatus in the SFR between the end of the halo/thick disk phase and the beginning of the thin disk phase. Such a hiatus could be real since it is observed both in the plot of [Fe/O] vs. [O/H] (Gratton et al. 1996, 2000) and in the plot of [Fe/Mg] vs. [Mg/H] (Fuhrmann 1998, Gratton et al. 2000). The evidence for this is shown by the steep increase of [Fe/O] and [Fe/Mg] at a particular value of [O/H] and [Mg/H], respectively, indicating that at the epoch of halo-disk transition Type II supernovae (O, Mg main producers) stopped exploding.

\subsection{Nucleosynthesis prescriptions}
The nucleosynthesis prescriptions adopted here are the same as in Romano et al. (2010; their model~15). These yields have proven to be the best to reproduce the [X/Fe] vs. [Fe/H] relations for several elements in the solar neighbourhood. The primordial abundances of D and \element[ ][3]{He} are $2.6\times10^{-5}$ and $0.9\times10^{-5}$ (by number with respect to hydrogen), respectively, while the adopted primordial abundance of \element[ ][4]{He} is 0.248 (by mass). These quantities are consistent with both predictions from the standard Big Bang nucleosynthesis theory and limits from observations of scarcely evolved systems.

\subsubsection{Low- and intermediate-mass stars}

Stars with initial masses between 1 and 5-8 $\mathrm{M}_{\odot}$ (depending on the overshooting adopted in stellar evolutionary models) pass through a phase of double shell burning, referred to as thermally-pulsing asymptotic giant branch (TP-AGB) phase, before ending their lives as white dwarfs.

Here we consider the AGB yields of Karakas (2010), calculated from detailed stellar models where the structure was computed first and the nucleosynthesis calculated afterwards using a post-processing algorithm. Yields are computed for 77 nuclei including all stable isotopes from H to \element[ ][34]{S} and for a small group of Fe-peak nuclei. The models cover a mass range 1.0-6 $\mathrm{M}_{\odot}$ and metallicity compositions $Z=0.02, 0.008, 0.004, 0.0001$. 

When occurring in binary systems, low- and intermediate-mass stars may experience thermonuclear runaways (nova systems) or even go disrupted in more dramatic explosive events (Type Ia supernovae). In order to take into account the chemical enrichment by Type Ia supernovae and nova systems, the yields of Iwamoto et al. (1999) and those of Jos\'e \& Hernanz (1998) have been used, respectively.

\begin{figure*}[t]
\centering
\includegraphics[scale=0.9]{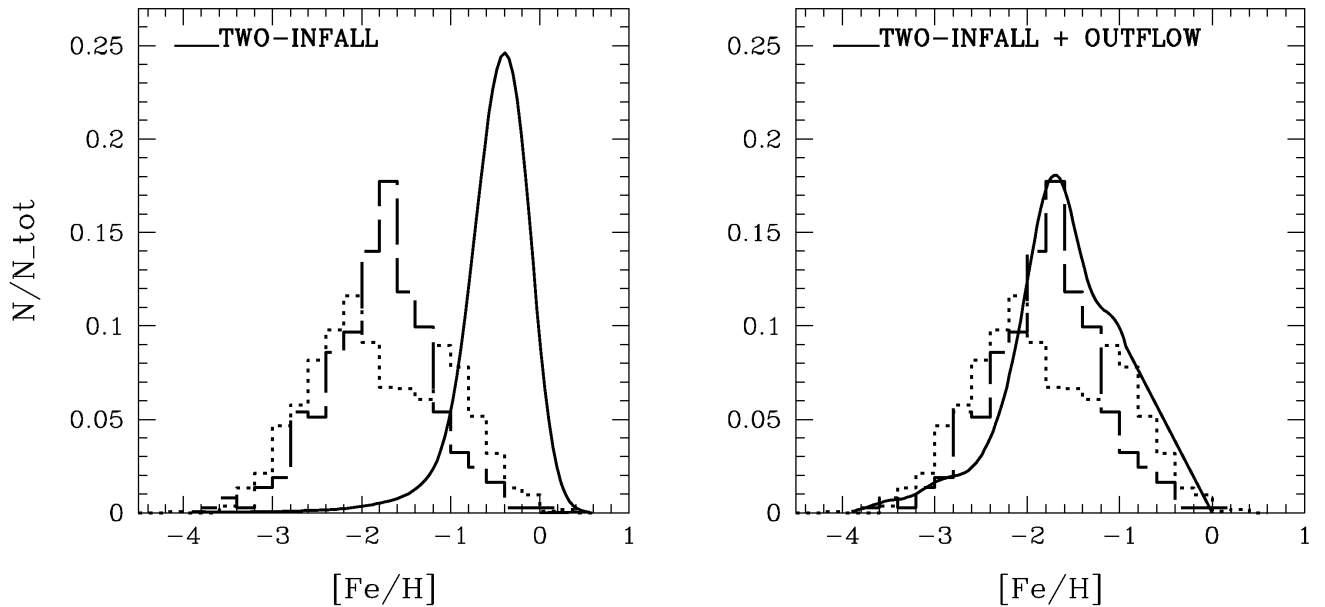}
\caption{Halo metallicity distribution function of the two-infall (left panel, solid) and two-infall + outflow model (right panel, solid) compared to the observational data from Ryan \& Norris (1991) (dashed) and Sch\"orck et al. (2009) (HES) (dotted). Theoretical predictions have been smoothed by a gaussian function with a variance equal to the data error, 0.2 dex.}
\label{fig:mdfwindow1}
\end{figure*}

\subsubsection{Massive stars}

Only a small fraction of the stellar mass, from 7 to 20\%, depending on the choice of the IMF, resides in stars with masses above 8 $\mathrm{M}_{\odot}$. Nevertheless, these stars are the major galactic polluters. They produce almost the whole body of the $\alpha$- and probably r-process elements and, if rotating fast, large amounts of primary \element[ ][13]{C}, \element[ ][14]{N}, \element[ ][22]{Ne} and s-process elements at low metallicities (see Meynet et al. 2009, for a review on rotating high-mass stars). It is now acknowledged that there exist two types of core-collapse supernovae. One is normal Type II supernovae (together with Ib and Ic), with explosion energies of the order of $10^{51}$ ergs. The other is hypernovae, with more than 10 times larger explosion energies (Iwamoto et al. 1998). Here, we adopt the yields of Kobayashi et al. (2006) who have calculated the yields for both Type II supernovae and hypernovae as functions of the initial masses (from 13 to 40 $\mathrm{M}_{\odot}$) and metallicities ($Z=0, 0.001, 0.004, 0.02$) of the progenitor stars. 

The inclusion of rotation in stellar evolutionary models considerably alters the outputs of traditional nucleosynthesis. Yields of \element[ ][4]{He}, \element[ ][12]{C}, \element[ ][14]{N} and \element[ ][16]{O} are available for high-mass rotating stars, for a fine grid of initial masses and metallicities (Meynet \& Maeder 2002a; Hirshi et al. 2005; Hirshi 2007; Ekstr\"om et al. 2008). Yields of \element[ ][13]{C}, \element[ ][17]{O} and \element[ ][18]{O} are provided only for a subset of models. 

The grid of yields for high-mass rotators that we use in this study has been obtained by patching together results presented in different papers by the same group (references are at the beginning of this paragraph).

\section{Two-infall + outflow model}

The outputs of the two-infall model allowed us to produce the halo MDF, which represents the number of halo stars with metallicity [Fe/H] in a certain range divided by the total number of halo stars as a function of the metallicity [Fe/H]. 
These predictions have been compared to the observational data of halo stars from Ryan \& Norris (1991) and Sch\"orck et al. (2009) (Hamburg/ESO Survey; HES): the model is not in agreement with the data, as shown in the left panel of Fig. \ref{fig:mdfwindow1}. In fact, the peak of the predicted MDF is located at about [Fe/H]~=$-0.3$, whereas the Ryan \& Norris (1991) data set peaks at [Fe/H]~$=-1.6$. This means that the model predicts too few stars at low metallicities with respect to what is observed in the Galactic halo. The theoretical distribution is not only displaced (by +1.3 dex) relative to the observed one, it is also much sharper.

This disagreement between predictions and observations led us to modify the model: adding a gas outflow during the halo formation phase has shown to be a good way to explain the halo MDF. In fact, in this case the metallicity of gas and stars is reduced because of the mass loss; this results in a higher number of predicted metal-poor stars with respect to the two-infall model without outflow, as the observations suggest.

Hartwick (1976) first suggested that, in the framework of the simple model of galactic chemical evolution, a mass outflow at a rate proportional to the SFR, $W(t)=\lambda\psi(t)$, must have occurred during the halo formation. He proposed the value of $\lambda=10$. This assumption is entirely plausible if supernovae and hot stars are responsible for causing the temporary removal of the gas from 
regions of active star formation. It is also possible that some of this gas was lost from the Galaxy completely in a galactic wind. 

More recently, Prantzos (2003) noticed that this \emph{`simple outflow model explained readily the peak and the overall shape of the halo MDF, but it failed to describe the lowest-metallicity part of it and the discrepancy is larger when the istantaneous recycling approximation is relaxed (as it should be)'}. He suggested that assuming a gas infall during the early phases produces a better agreement between theory and observations concerning the very low metallicity part of the MDF. In fact, in the pure outflow model the system had to `wait' until the total amount of gas became available to start the star formation and then eject (part of) that gas. Instead, it is more reasonable that stars formed since the beginning of the Galaxy formation while the system was still accreting gas. 

In the two-infall + outflow model, we assume that at the beginning of the Galaxy formation part of the collapsing gas forming halo stars continued to collapse towards the center, thus producing an outflow from the halo component. However, we do not exclude that part of this gas can be lost from the Galaxy.

The outflowing wind has been inserted in the form \`a la Hartwick (1976) with a rate proportional to the SFR. 
In particular, the outflow rate is defined as:

\begin{equation}
\frac{d\sigma_{W}}{dt}=- \lambda \psi(t)
\end{equation}
where the SFR is given in eq. (2).

The best agreement with observational data (see next Section) has been obtained modifying also the parameter $\tau_H$, the time scale for the mass accretion during the halo formation. The parameters have been best set to $\lambda=14$ and $\tau_H=0.2$ Gyr.

\begin{figure}[t]
\centering
\includegraphics[scale=0.4]{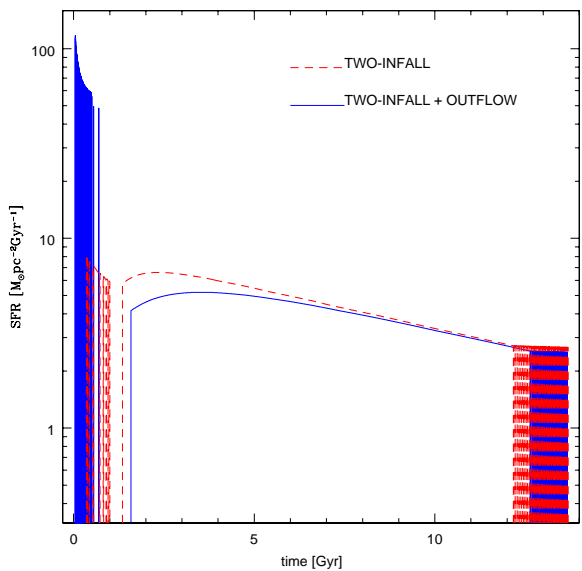}
\caption{Star formation rate as a function of time predicted by the two-infall model (red dashed) and by the two-infall + outflow model (blue solid). The oscillating behaviour during both the halo and the last disk phases is an effect of the gas density threshold for star formation (see Chiappini et al. 1997).}
\label{fig:sfrboth}
\end{figure}

\begin{figure*}[t]
\centering
\includegraphics[scale=0.9]{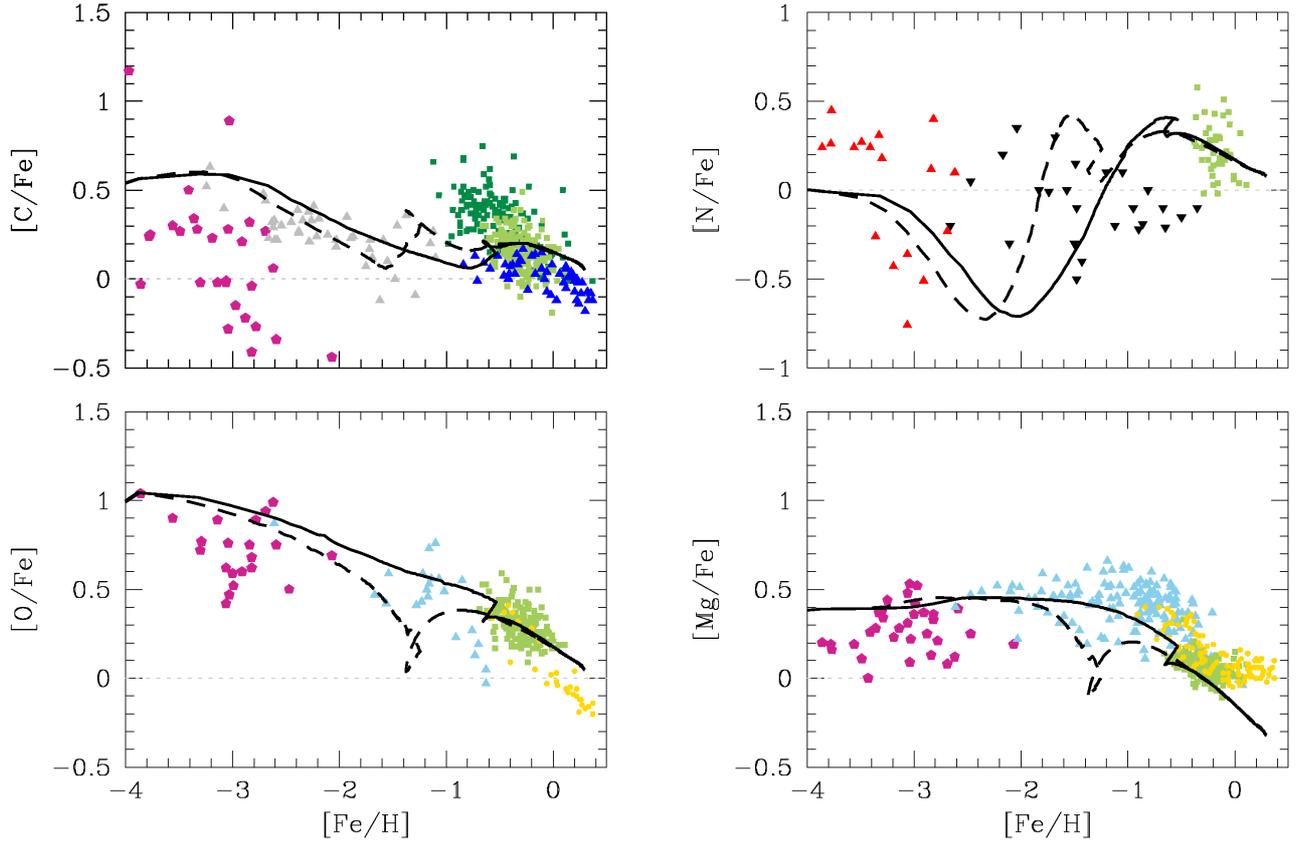}
\caption{Abundance patterns [X/Fe] vs. [Fe/H] for carbon, nitrogen, oxygen and magnesium as predicted by the two-infall model (solid) and by the two-infall + outflow model (dashed). The adopted solar abundances are those of Asplund et al. (2009). Data are from Cayrel et al. (2004) (fuchsia pentagons), Fabbian et al. (2009) (grey triangles), Reddy et al. (2006) (dark-green squares), Reddy et al. (2003) (light-green squares), Bensby \& Felzing (2006) (blue triangles), Spite et al. (2005) (red triangles), Israelian et al. (2004) (black upside-down triangles), Gratton et al. 2003 (light-blue triangles), Bensby et al. 2005 (gold circles). At very low metallicities nitrogen is produced as a primary element, according to the Geneva yields including stellar rotation. These yields allow us to explain data at very low metallicities, but still we do not reproduce [N/Fe] between -3 and -2 dex. Primary nitrogen from all massive stars (not just highly rotating) would be necessary.}
\label{fig:abundwindow1}
\end{figure*}

\begin{figure*}[t]
\centering
\includegraphics[scale=0.9]{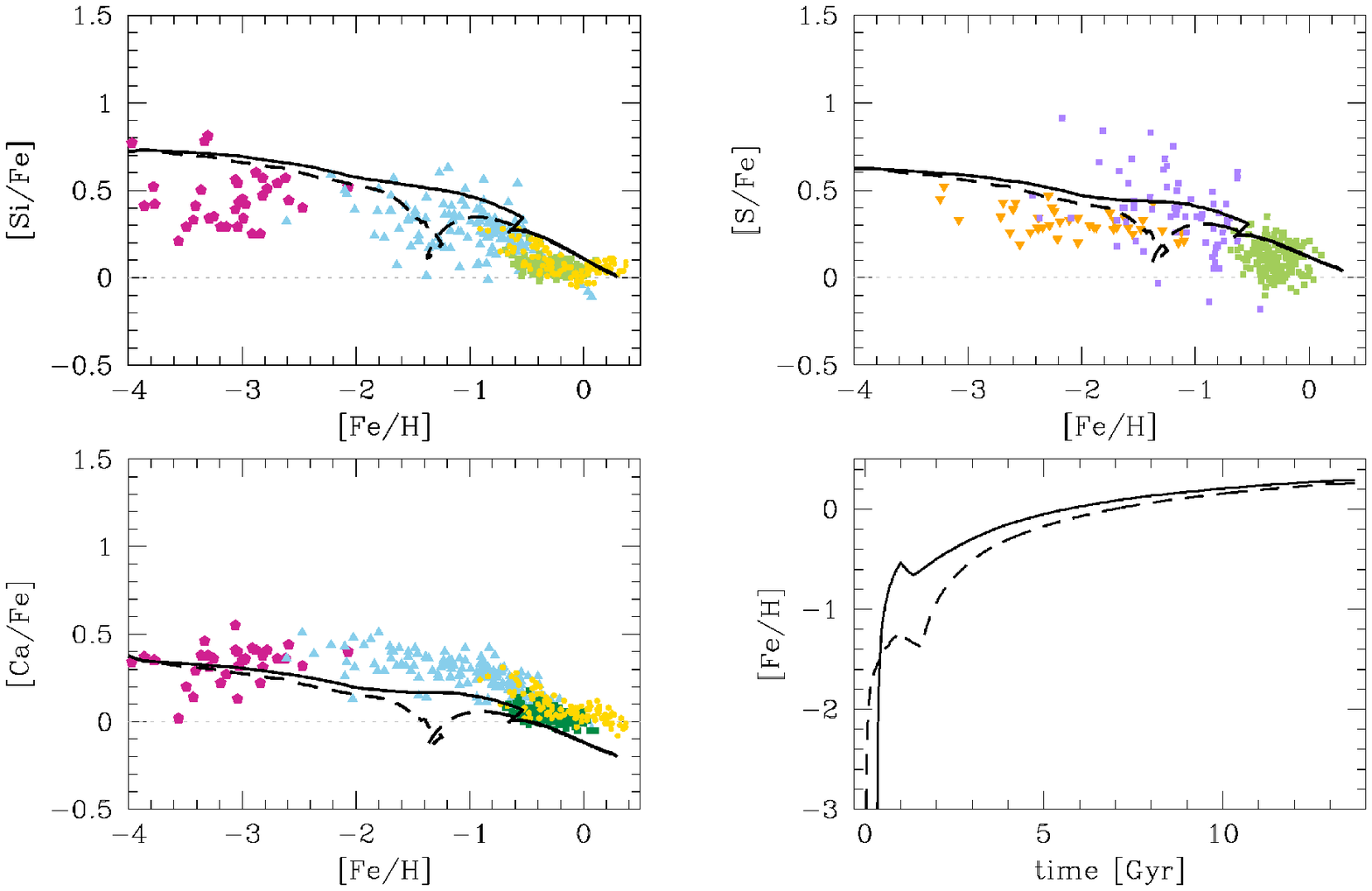}
\caption{Abundance patterns [X/Fe] vs. [Fe/H] for silicon, sulfur and calcium and [Fe/H] abundance as a function of time (lower right panel) as predicted by the two-infall model (solid) and by the two-infall + outflow model (dashed). The solar abundances have been updated to those of Asplund et al. (2009). Data as in Fig. \ref{fig:abundwindow1} and from Nissen et al. (2007) (orange upside-down triangles) and Caffau et al. (2005) (purple squares).}
\label{fig:abundwindow2}
\end{figure*}

\subsection{Results}

The resulting MDF is presented in the right panel of Fig. \ref{fig:mdfwindow1}, which shows the good agreement between predictions and observations produced by the addiction of the outflow. Hence, we confirm that it is fundamental to include both a gas infall and outflow during the halo formation in modeling the chemical evolution of the Galaxy, to explain the observed MDF of halo stars (see also Prantzos, 2003). Moreover, the value of the parameter $\lambda=14$ is the same suggested by Cescutti \& Chiappini (2010) in their inhomogeneous simple model for the Galactic halo. Further, we obtain that the halo formation happened on a time scale ($\tau_H=0.2$ Gyr) much shorter than the age of the Galaxy ($\sim$14 Gyr), suggesting a fast formation for the halo. The new parameter $\tau_H$ is not very different from the one used previously in the two-infall model ($\tau_H=0.8$ Gyr) and both models assume that $t_{max}$, the time of maximum gas accretion onto the disk, coincident with the end of the halo/thick disk phase, is equal to 1 Gyr. 

The insertion of an outflow term changes also the SFR during the halo phase, as presented in Fig. \ref{fig:sfrboth}. Since in the two-infall + outflow model the infall timescale is smaller than in the two-infall model, more gas is present at early times and the SFR is higher, as the figure shows. Another interesting aspect is that in the two-infall + outflow model the SFR becomes zero (because of the gas density threshold for star formation) immediately after the halo formation phase for a time interval of $\sim$1 Gyr, whereas in the case of the two-infall model without outflow this time interval was less than 0.5 Gyr. This is due to the fact that, since mass is lost in the outflowing wind, the system needs to wait longer until the infalling mass reaches the gas density threshold to restore star formation.

Another observational constraint for our predictions is represented by the abundance pattern [X/Fe] vs. [Fe/H] for various elements X. Here we show the results for C, N and $\alpha$-elements O, Mg, Si, S, Ca (Figs. \ref{fig:abundwindow1} and \ref{fig:abundwindow2}). 

The agreement of the two-infall + outflow model with the observational data is good, as it was for the two-infall model  (see Romano et al. 2010, their figure~22). In fact, from the outflow we do not expect any consistent effect on abundance ratios, since the wind acts in the same way on each element. 

 Nevertheless, we notice that in the case of outflow the `knot'-like feature of the predicted patterns is shifted towards lower metallicities with respect to the classical two-infall case. This `knot' is an effect of the adopted threshold for the star formation: if the star formation is inactive, some elements, such as O, stop to be produced whereas others, such as Fe, continue to be produced, thus forming the `knot' (see also Chiappini et al. 1997). The effect of the outflow on the abundance patterns can be clearly seen in the plot of [Fe/H] vs. time for the two models, shown in the lower-right panel of Fig. \ref{fig:abundwindow2}: the `knot' is shifted towards lower metallicities and it extends for a longer time interval. We explain this considering that in the two-infall + outflow model the gap in the SFR is longer than in the case without outflow (Fig. \ref{fig:sfrboth}), thus producing different abundance patterns. This explanation is confimed considering that in Figs. \ref{fig:abundwindow1} and \ref{fig:abundwindow2} the [Fe/H] values delimiting the `knot' correspond, through the age-metallicity relation, to the extremes of the time interval in which the SFR is zero.  

The abundance patterns for $\alpha$-elements are well interpreted in the framework of the time-delay model: `time-delay' refers to the delay of iron ejection by Type Ia supernovae relative to the faster production of $\alpha$-elements by core-collapse supernovae. The effect of the delayed iron production is to create an overabundance of $\alpha$-elements relative to iron ([$\alpha$/Fe]$>0$) at low [Fe/H] values, and a continuous decline for [Fe/H]$>-1.0$ dex of the [$\alpha$/Fe] ratio until the solar value ([$\alpha$/Fe]$_{\odot}=0$) is reached. In addition, the stellar yields of nitrogen adopted here are strongly dependent on the rotational velocities of metal-poor stars, so it is possible to understand the apparent observational contradiction consisting in a large scatter in [N/Fe] and an almost complete lack of scatter in [$\alpha$/Fe] ratios found in the same very metal-poor halo stars. Although the observed scatter could be related to the distribution of stellar rotational velocities as a function of metallicity, that the neutron-capture elements in the same stars also show a large scatter pointed to a strong variation in the stellar yields with the mass range of the stars responsible of the synthesis of these elements (Cescutti \& Chiappini 2010). Cescutti (2008) explains simultaneously the observed spread in the neutron capture elements and the lack of scatter in the $\alpha$-elements as being caused by the stochasticity in the formation of massive stars, combined with the fact that massive stars belonging to different mass ranges are responsible for the synthesis of different chemical elements, namely: only massive stars with masses between 12 and 30 $\mathrm{M}_{\odot}$ contribute to the neutron capture elements, whereas the whole mass range of massive stars (10 to 100 $\mathrm{M}_{\odot}$) contribute to the production af $\alpha$-elements.
Finally, we would like to mention that it has been often suggested, in the framework of the hierachical formation of galaxies, that the Galactic halo stars could have been accreted from dwarf satellites. Abundance ratios, such as [$\alpha$/Fe], have been measured both in halo and dwarf stars. It has been found that there is a similarity in these ratios only for very metal poor stars in dwarf galaxies, but that the overall pattern of the [$\alpha$/Fe] vs. [Fe/H] is different indicating a faster decrease of the [$\alpha$/Fe] ratios with [Fe/H] in the satellite galaxies.
It has been shown by Lanfranchi \& Matteucci (2004) that the  [$\alpha$/Fe] ratios in dwarf spheroidals can be well reproduced by models assuming a SFR much slower that in the Galactic halo coupled with a strong galactic wind. In this case, the SF histories of the Milky Way halo and dwarf satellites have been very different, but we cannot exclude that few of the first stars in dwarf satellites might have contributed to form the halo.

\section{Population III stars} 

It is generally believed that a population of massive stars with zero-metal content existed at the beginning of the Galaxy formation. We test this hypothesis on the two-infall + outflow model, by adopting two different top-heavy initial mass functions and the population III yields of Heger \& Woosley (2002) to compare the resulting MDFs with observational data.
Zero-metal first stars have not been observed yet and the primordial IMF represents one of the most discussed topics in this field. Theoretical studies support the idea of a top-heavy IMF, biased towards massive and very massive stars (e.g. Bromm et al. 2002). In fact, in the absence of metals, the collapsing gas fragments in $\sim$$10^3$ $\mathrm{M}_{\odot}$ clumps, progenitors of the protostars which will later accrete gas on the central core and form stars. The final stellar masses are still largely uncertain but likely to be in the range 30-300 $\mathrm{M}_{\odot}$ (Tan \& McKee 2004). However, recent simulations (e.g. Stacy et al. 2012) have revised the Pop III mass-scale downwards to more modest values (10-50~$\mathrm{M}_{\odot}$).
In contrast, present-day star formation (pop II/I stars) is characterized by a Salpeter-like IMF in the mass range 0.1-100 $\mathrm{M}_{\odot}$, flattening below masses of 0.35 $\mathrm{M}_{\odot}$ (Larson 1998) and the characteristic mass of local pop II/I stars is $\sim$1 $\mathrm{M}_{\odot}$. Recent theoretical studies suggest that there must have been a transition in the properties of star-forming regions through cosmic times, in particular the initial metallicity of the star-forming gas represents the key element controlling this transition. Schneider et al. (2002) found that when the metallicity is in the critical range $10^{-6}<Z_{cr}/Z_{\odot}<10^{-4}$ there occurs a transition in fragmentation scales from $\sim$$10^3$ $\mathrm{M}_{\odot}$ to solar or subsolar fragments, thus producing a transition in characteristic stellar masses. Therefore, according to this scenario, when the metallicity exceeds the value of $Z_{cr}$, the presence of metals triggers the onset of low-mass star formation in the Universe.
Massive population III stars should start their main-sequence phase burning hydrogen by means of the CNO reaction chain; but since they lack CNO elements, they are forced to burn hydrogen via the p-p chain. However, the p-p chain is not so effective in producing the energy necessary to counterbalance their gravity. Therefore, these stars are forced to contract and increase their central temperature until the 3$\alpha$ reaction starts and creates some \element[ ][12]{C} and O; at this point, they can start burning hydrogen via the CNO-cycle. Because of the lack of metals, these stars suffer less mass loss, due to radiation pressure where metals represent the main photon absorbers, than stars belonging to populations II and I. As a consequence, they end their lives by creating a black hole in the center and explode as pair-creation supernovae, which leave no remnant. Clearly the nucleosynthesis products of population III stars are different from those of pop II/I stars. 

\begin{figure*}[t]
\centering
\includegraphics[scale=0.9]{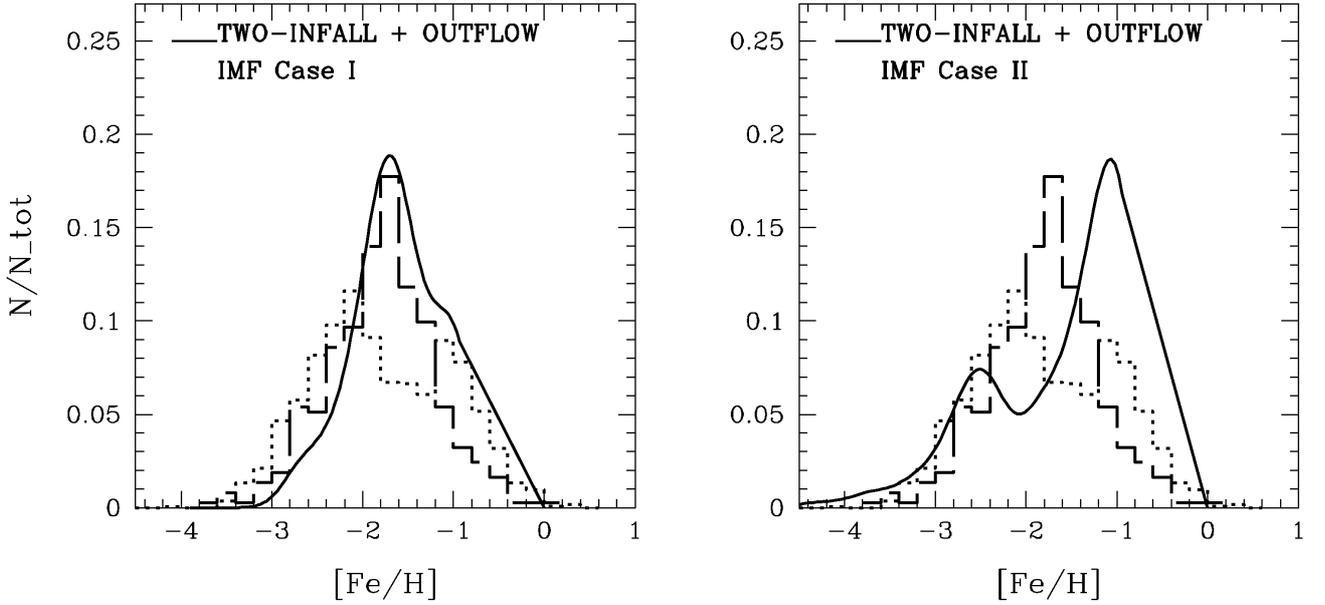}
\caption{Halo metallicity distribution function of the two-infall + outflow model with case I IMF (left panel, solid) and case II IMF (right panel, solid). Observational data are from Ryan \& Norris (1991) (dashed) and Sch\"orck et al. (2009) (HES) (dotted). The theoretical prediction has been smoothed by a gaussian function with a variance equal to the data error, 0.2 dex.}
\label{fig:mdfwindow2}
\end{figure*}

\subsection{Nucleosynthesis in pop III stars}

Heger and Woosley (2002), among others, explored the nucleosynthesis of helium cores on the mass range $M_{He}=64$-133 $\mathrm{M}_{\odot}$, corresponding to main-sequence star masses of approximately 140-260 $\mathrm{M}_{\odot}$ through the `mass-mass of the helium core' relation $M_{He}\approx\frac{13}{24}(M_{ZAMS}-20$ $\mathrm{M}_{\odot})$. They found that above $M_{He}=133$ $\mathrm{M}_{\odot}$ a black hole is formed and no chemical elements are ejected. For lighter helium core masses, $\sim$40-63 $\mathrm{M}_{\odot}$, violent pulsations occur, induced by the pair instability and accompanied by supernova-like mass ejection, but the star eventually produces a large iron core in hydrostatic equilibrium and it is likely that this core, too, collapses to a black hole. \emph{`Thus'}, Heger \& Woosley concluded, \emph{`the heavy-element nucleosynthesis of pair creation supernovae is cleanly separated from those of other masses. Indeed, black hole formation is a likely outcome for all population III stars with main sequence masses between about 25 and 140 $\mathrm{M}_{\odot}$ ($M_{He}=9$-63 $\mathrm{M}_{\odot}$) as well as those above 260 $\mathrm{M}_{\odot}$'}. 

\subsection{Two top-heavy IMFs}

In order to test the effects of a possible stellar population III in the chemical evolution of the halo, we considered an initial top-heavy IMF and adopted the yields of Heger \& Woosley (2002) for stars with mass $140<M/\mathrm{M}_{\odot}<260$. Assuming the critical metallicity to be equal to $Z_{cr}=10^{-4}$~$Z_{\odot}$ ($Z_{\odot}=0.0134$, Asplund et al. 2009), we studied two cases:

\begin{itemize}
\item Case I:

\[
\Phi(m)\propto m^{-1-x}
\]
where
\[
\left\{
\begin{array}{lcr}
\mathrm{if}~ Z\le Z_{crit}~\mathrm{and} & 10<m/\mathrm{M}_{\odot}<260: & x=1.7;  \\
\mathrm{if}~ Z>Z_{crit}~\mathrm{and}  & 0.1\le m/\mathrm{M}_{\odot}<0.5: & x=0.3, \\
		 		    				   & 0.5\le m/\mathrm{M}_{\odot}<1.0:  & x=1.2, \\
			               			  & 1.0\le m/\mathrm{M}_{\odot}\le100: & x=1.7. 		
\end{array}
\right.
\]

\item Case II:

\[
\Phi(m)\propto m^{-1-x}e^{-\frac{m_{cut}}{m}}
\]
where
\[
\left\{
\begin{array}{lcr}
\mathrm{if}~Z\le Z_{crit}: & x=1.35, m_{cut}=10 \mathrm{M}_{\odot};  \\
\mathrm{if}~Z>Z_{crit}: & x=1.35, m_{cut}=0.35 \mathrm{M}_{\odot}.
\end{array}
\right.
\]

\end{itemize}

In case~I, it is assumed that, below $Z_{cr} = 10^{-4}$~$Z_{\odot}$, the minimum stellar mass was 10 $\mathrm{M}_{\odot}$. The adopted IMF is a one-slope power law until the critical metallicity is reached, while for higher metallicities it is the Kroupa (1993) IMF (see also Ballero et al. 2006). In case~II, instead, the 
IMF is similar to the one used in Salvadori et al. (2007).

\subsection{Results for PopIII stars}

The resulting MDFs are shown in Fig. \ref{fig:mdfwindow2}. A first generation of population~III stars is introduced by adopting the SFR in the early 
phases which is shown in Fig. 2 (blue solid line). The SF starts, in our model, only after 0.26 Gyr from the beginning of the gas accretion into the dark matter halo of the Galaxy (first infall episode). This is due to the fact that we assume a threshold gas density for star formation and stars do not form before such a threshold is reached. As a consequence, the first stars are formed at redshift $z\sim 15$, in agreement with cosmological simulations (e.g. Maio et al. 2010; Wise et al. 2012). Looking at Fig. \ref{fig:mdfwindow2} we can see that the overall agreement of the model predictions of the two-infall~+~outflow model with the observational data worsens relative to the case with no PopIII stars. In particular, in case I the model does not reproduce the very low-metallicity tail of the observed MDF anymore. No low mass stars with [Fe/H]$< -3.2$ dex are found.
In this case, the critical metallicity $Z_{cr}=10^{-4}Z_{\odot}$ is reached in $\sim$ 3 Myr after the beginning of star formation, and it is due to the very short lifetimes of the pair creation SNe (140-260 M$_{\odot}$). The total mass of PopIII stars necessary to reach this critical metallicity is a fraction of $\sim 4.3 \cdot 10^{-4}$ of all the halo stars, while the mass of gas polluted by the PopIII stars is $\sim 0.075$ of the stellar halo mass which is $\sim 5 \cdot 10^{8} M_{\odot}$.
It is worth noting that the timescale for pollution of PopIII stars we find is clearly a rough estimate and a lower limit: in fact, we assume instantaneous mixing approximation and relaxing this assumption could imply a longer timescale for pollution of the whole halo. However, cosmological simulations (e.g. Maio et al. 2010; Wise et al. 2012), which account for mergers and inhomogeneities, have indicated a very similar timescale, of the order of a few millions years, and with a maximum of $2\cdot 10^{7}$ years. On the other hand, we recall that Trenti \& Shull (2010), by means of cosmological simulations, give a value $<10^{8}$ years even considering very massive stars, but this long timescale has not been confirmed by subsequent studies.
It is worth noting that if the maximum mass for the PopIII stars is as low as 100~M$_{\odot}$, then we find that the timescale for reaching $Z_{cr}$ would be slightly longer (5-10 Myr). However, also in this case there would be no agreement with the observed MDF of the halo. Therefore, given the small critical metallicity, which could be even lower ($10^{-6}Z_{\odot}$) if cooling by dust is effective (e.g. Schneider et al. 2006), we are forced to conclude that the PopIII phase was extremely rapid, especially if masses as high as 260~M$_{\odot}$ are considered. 

The MDF predicted in case II is even more unsatisfying: while at very low metallicities the case II predictions are in agreement with the observational data, the overall halo MDF is poorly fitted by this model. It must be said that Salvadori et al. (2007) adopted this particular IMF to study the Galactic halo but their comparison was focused only on the MDF for [Fe/H]~$<-2.0$ dex.

\section{Conclusions and discussion}

In this work, we studied the chemical evolution and formation of the Galactic halo, starting from the predictions of the numerical two-infall model developed first in Chiappini et al. (1997) and updated by Romano et al. (2010). In order to better compare the model results to the observational data of halo stars, we included  a gas outflow during the halo phase. In particular, one of the most powerful probes we tested is the MDF of halo stars, together with some element abundance patterns in the Galaxy. We referred to the observational halo MDF data sets of Ryan \& Norris (1991) and Sch\"orck et al. (2009) (HES). We modified the two-infall model because it predicts too few stars at low metallicities with respect to what is observed in the Galactic halo. Adding a gas outflow during the halo phase helps in explaining the observed halo MDF. In fact, in this case the metallicity of gas and stars is reduced because of the mass loss; this results in a higher number of predicted metal-poor stars with respect to the two-infall model without outflow, as observations suggest. The outflowing wind has been inserted in the form \`a la Hartwick (1976), with a rate proportional to the SFR, 
$W(t)=\lambda\psi(t)$. The best agreement with the observational data 
is obtained in this context by modifying also the parameter $\tau_H$, the time scale for the mass accretion during the halo formation. The parameters have been best set to $\lambda=14$ and $\tau_H=0.2$ Gyr (rather than 0.8~Gyr as in the model without outflow). Hence, in the presence of an outflow, the evolution of the halo must have been more rapid.

In addition, we tested the existence of population III stars (zero-metal content) at the beginning of the Galaxy formation on the two-infall + outflow model. The first stars form out of gas of primordial composition, so they are expected to be metal free. They release metals in the interstellar medium, that eventually enrich the intergalactic medium through supernova winds; out of this enriched gas, population II stars (including halo stars) form. Pop III stars have not been observed yet and the primordial IMF represents one of the most discussed topics in this field. Theoretical studies support the idea of a top-heavy IMF, biased towards massive and very massive stars. Therefore, in the two-infall + outflow model we adopted two different top-heavy initial mass functions, until the critical metallicity $Z_{cr}=10^{-4}$~$Z_{\odot}$ ($Z_{\odot}=0.0134$) is reached, and the population III yields of Heger \& Woosley (2002) to compare the resulting MDFs with observational data.

Our conclusions are:

\begin{itemize}

\item It is fundamental to include both a gas infall and outflow during the halo formation in modeling the chemical evolution of the Galaxy, to explain the observed MDF of halo stars. Our best value for the outflow parameter is $\lambda=14$. The abundance patterns [X/Fe] vs. [Fe/H] for C, N and $\alpha$-elements (O, Mg, Si, S, Ca) predicted by the two-infall + outflow model well reproduce the observational data, as in the case of  the two-infall model without outflow, both for the halo and disk stars. This is because, since the outflow is acting in the same way on different elements, its effects cancel out when looking at abundance ratios. Yet, the addition of the outflow results in a duration of the gap in the SFR between the halo/thick disk and thin disk phases longer than predicted by Chiappini et al. (2001): this longer duration (1 Gyr) has the effect of shifting the `knot'-like feature in the [X/Fe] vs. [Fe/H] pattern towards lower metallicities relative to the standard two-infall case.

\item After testing two different top-heavy population III IMFs on the two-infall + outflow model, we can say that the overall agreement of the predicted MDFs with observational data is worse with respect to the case of the two-infall + outflow model without population III. In particular, adopting a one-slope ($x=1.7$) IMF with a low-mass cut at 10~M$_{\odot}$ and a high-mass cut at 260~M$_{\odot}$ until the critical metallicity is reached, and the classic three-slope Kroupa et al. (1993) IMF for higher metallicities, the model predicts too few low mass stars at low metallicities and clearly it cannot explain the star recently discovered by Caffau et al. (2011) with an estimated global metal content 
$Z<4.5 \cdot 10^{-5}$~$Z_{\odot}$, lower than $Z_{cr}$. This, together with the fact that we predict that  $Z_{cr}$ is reached in a timescale of $\sim$ 3 Myr, suggests that only very few population III stars must have existed.
 
The MDF predicted in the case with a one slope IMF ($x=1.35$) for the entire Galactic lifetime, but with a variable lower mass cut-off, is even more unsatisfying: while at very low metallicities the predictions are in agreement with the observational data, the overall halo MDF is poorly fitted by this model.

We conclude that it does not exist, yet, a satisfactory form for the IMF, including population III stars, that well explains the observed halo MDF; we find that the best fit of the observed halo MDF remains that predicted by the two-infall + outflow model without population III stars. 

\end{itemize}

\subsection{Discussion}
Prantzos (2008) was able to fit the halo stellar MDF by assuming that the halo of the Milky Way could be composed by a number of sub-units with stellar masses ranging from $2\times 10^6$~M$_{\odot}$ to $2\times 10^8$~M$_{\odot}$. These sub-units should be similar (but not identical) to Local Group dwarf galaxies. 
A formation scenario for the halo from fragments would be consistent with the requirement that present-day Galactic globular clusters have had more massive progenitors (see, e.g., Gratton et al . 2012, and references therein). We note that in our approach, mergers with smaller systems are indeed allowed, provided the accreted proto-Galactic clumps are mostly gaseous and metal-poor. Also, it should be stressed that our proposed formation scenario applies to the inner Galactic halo (indeed, the HES predominantly samples the inner-halo population of the Galaxy, $R < 15$ kpc), which is likely to have formed differently from the outer halo component (Carollo et al. 2007).
Concerning the question whether the halo stars have formed {\it in situ} or have been accreted from dwarf satellites, there is one observational fact to be taken into account: the generally different shapes of the [$\alpha$/Fe] vs. [Fe/H] relation in the halo and dwarf spheroidals. On the other hand, in this paper we have shown that the hypothesis that the halo stars formed in situ can well explain their MDF and abundance ratios.

In conclusion, a robust solution to the problem of the halo formation has still to be found. In the coming years, our knowledge of the halo metallicity distribution function is likely to improve significantly with surveys such as Skymapper, SEGUE and LAMOST, which will extend by large factors the currently known handful of extremely metal-poor stars. These new data sets will give us access to the earliest phases of the Universe through the fossil record left by the first stars formed (Helmi 2008). In the meanwhile, the study of substructures within the Galaxy (see, e.g., Majewski et al. 2003; Belokurov et al. 2006; Johnston et al. 2012) will continue to provide precious hints on the frequency and significance of merging events in building up the different Galactic components. 

\begin{acknowledgements}
G.B. thanks Gabriele Cescutti for useful suggestions. F.M. and D.R. aknowledge financial support from PRIN MIUR 2010-2011, project ``The Chemical and dynamical Evolution of the Milky Way and Local Group Galaxies'', prot. 2010LY5N2T. F.M. thanks U. Maio for very useful discussions.
Finally, we also like to thank an anonymous referee for carefully reading the paper and giving suggestions which improved the paper.
\end{acknowledgements}

\end{document}